\documentclass[12pt,a4paper]{article}
\usepackage[cp1251]{inputenc}
\usepackage[russian,english]{babel}
\usepackage{graphicx}
\usepackage{amsmath}
\usepackage{amssymb}

\lccode`\-=`\-
\date{}

\large
\title{Coherent Schwinger Interaction
from Darboux Transformation}
\author{Ekaterina Pozdeeva\\Semenov Institute of Chemical Physics,\\
 Russian Academy of Sciences \footnote{E-mail: ekatpozdeeva@mail.ru}}
 
\begin{document}
\selectlanguage{english} \maketitle
\begin{abstract}
The exactly solvable scalar–tensor potential of the four-component
Dirac equation has been obtained by the Darboux transformation method.
The constructed potential has been interpreted in terms of
nucleon–nuclear and Schwinger interactions of neutral particles with
lattice sites during their channeling in the nonmagnetic crystal. The
family of exactly solvable interaction Hamiltonians of a Schwinger type
is obtained by means of the Darboux transformation chain. The analytic
structure of the Lyapunov function of periodic continuation for each of
the Hamiltonians of the family is considered.
\end{abstract}

\section{Introduction}
The Darboux (Moutard) transformation \cite{Darbu,Moutard,Rosu} is
widely used for constructing solvable quantum mechanic models
\cite{Abl,Matv,BAGROVECHAR,Pupasov}. It is  known as the method of
construction of the  solvable potentials for the one-dimensional
Schr\"o\-din\-ger equation \cite{Sukumar1985S,BagrovSamsonov,per2,
Suzko} and one for two-component case was considered in
\cite{SUZKO}. The quaternionic factorization of the
Schr\"o\-din\-ger equation was obtained in \cite{Kravchenko}.

As was shown  in  \cite{1,2,3,4,5,7}, the exactly solvable models
for the one-dimensional Dirac equation with the Lorentz scalar
potential employ  the close relation between this equation and a
pair of the Schr\"o\-din\-ger equations. The solution of these
equations, in turn, is  based on well established algebraic
techniques. In papers
\cite{Stalhofen,annphys2003v305p151,Eurjphys,We} the Darboux
transformation was applied directly to the one-dimensional
two-component Dirac equation, where as for applications it is often
important the four-component Dirac equation \cite{prosaiding,Bord}.

In works \cite{Anderson,Yurov} the attempts to use the Darboux
transformation to the four-component Dirac equation have been done.
However, as a result, the Darboux transformation was applied to
two-component equations.

 In \cite{Kravchenko} the Darboux
transformation of the Dirac equation was reduced to the quaternionic
factorization of the Schr\"o\-din\-ger equation. The Darboux---Crum
method  \cite{Darbu,Crum} was applied to the Schr\"o\-din\-ger
equation in \cite{Andrianov} and to the two-component Dirac equation
in \cite{annphys2003v305p151}.

In the present  work  we apply the Darboux transformation to the
one-dimensional four-component Dirac equation,  thus we make
contribution for applying the Darboux transformation to the full
Dirac equation. Also we construct a chain of the Darboux---Crum
transformations for the four-component Dirac equation. We try to
generalize these results to construct an exactly solvable
interaction Hamiltonian, which allows one to solve analytically  the
one-dimensional four-com\-po\-nent Dirac equation.

The structure of the present paper is as follows. In section 2 we
construct a Darboux transformation for the free Hamiltonian of the
one-dimensional four-component Dirac equation and generate an
exactly solvable transparent scalar-tensor potential. We show that
the tensor part of the external potential has the meaning of the
electrical strength. In section 3 we consider a some special choice
of transformation parameters. In section 4 we construct a chain of
the Darboux---Crum transformations for the four-component Dirac
equation. In section 5 we briefly discuss the structure of the
Lyapunov functions for the problem of the interaction with the
periodic potentials. Finally, in section 6 we interpret the
interaction of a spin-$1/2$ neutral massive particle with an
external electrical field through its magnetic moment as coherent
Schwinger interaction and try to apply these results  to neutral
particles channeling in crystals.

\section{Darboux transformation of
 the one-dimensional four-component Dirac equation}
The stationary one-dimensional Dirac equation reads:
\begin{eqnarray}
  \label{1}H\psi=E\psi,\qquad H=H_0+H_{I}, \\
  H_0=-i\alpha_1\partial_x+\beta m,\qquad
  H_{I}=H_{I}(x)\nonumber,
\end{eqnarray}
where  $\alpha_1=\left(%
\begin{array}{cc}
  0 & \sigma_1 \\
  \sigma_1 & 0 \\
\end{array}%
\right),$
$\beta=\left(%
\begin{array}{cc}
1_2 &   0 \\
0 & -1_2\\
\end{array}%
\right),$
$1_2=\left(%
\begin{array}{cc}
  1 & 0 \\
  0 & 1 \\
\end{array}%
\right);$
 $\sigma_1$ is the Pauli matrix;  $H_0$ is the free Dirac Hamiltonian
\cite{Anderson,Yurov,Pozdeeva,thaler} and $H_{I}$ is the part of the
Hamiltonian describing an interaction of the spin-$1/2$ particle
with some external field. The solution of the equation (\ref{1})
 in the absence of the interaction is simple and well known for any
 $E$.

 Let us consider the Darboux transformation
\begin{eqnarray}
 \label{2} HL=LH_0, \qquad \psi=L\psi_0, \qquad H\psi=E\psi
  \end{eqnarray}
with the operator $L$ of the form
\begin{eqnarray}
\label{3} L=\partial_x-u_x u^{-1}.
\end{eqnarray}
Here $4\times 4$-matrix $u$ obeys the equation
\begin{eqnarray}
\label{4}
  H_0u=u\Lambda,
\end{eqnarray}
the $\Lambda$ is any nonsingular matrix,  $u_x=du/dx$.\\ Such
transformation generates the interaction  Hamiltonian
\begin{eqnarray}
\label{5}
  H_{I}=H-H_0=-i(\alpha_1u_xu^{-1}-u_xu^{-1}\alpha_1).
\end{eqnarray}
If we   verify that the function  (\ref{2}) is the solution of Eq.
(\ref{1}), where  $H_{I}$ is defined by (\ref{5}), we see hat  $u$
is definite by (\ref{4}).\\
 We choose
$\Lambda$ to be diagonal matrix of the form
\begin{eqnarray}
\label{6}
  \Lambda=[\lambda_1(I+\beta)+\lambda_2(I-\beta)]/2, \qquad
  \lambda_{1,2}\leq m.
\end{eqnarray}
Here $I=\left(%
\begin{array}{cc}
  1_2 & 0 \\
  0 & 1_2\\
\end{array}%
\right)$.\\
Let us construct the matrix $u$:
\begin{eqnarray}
\label{7}
  u=aI+b\alpha_1+c\beta+d\gamma, \qquad \gamma=\alpha_1\beta,
\end{eqnarray}
where
\begin{eqnarray}
\label{8} a=(\mu_1+\mu_3)/2,\quad b=(\mu_2+\mu_4)/2,\quad
 c=(\mu_1-\mu_3)/2,\quad d=(\mu_4-\mu_2)/2,
\end{eqnarray}
\begin{eqnarray}
\label{10}
 & &\mu_1=\cosh{(k_1x+\phi_1)},\qquad
\mu_3=\cosh{(k_2x+\phi_2)},\\
\label{11}&
&\mu_2=\frac{-ik_2\sinh{(k_2x+\phi_2)}}{\lambda_2-m},\qquad
\mu_4=\frac{-ik_1\sinh{(k_1x+\phi_1)}}{\lambda_1+m},
\end{eqnarray}
$k_{1,2}=\pm\sqrt{m^2-\lambda_{1,2}^2};$  $\phi_{1,2}$ are arbitrary
complex numbers. It is easy to check that these matrix
 is solution of the Eq.
(\ref{4}).\\
The inverse matrix reads:
\begin{eqnarray}
\label{9}
  u^{-1}=[aI-b\alpha_1-c\beta-d\gamma]/D,\qquad D=a^2-b^2-c^2+d^2.
\end{eqnarray}
With the help of simple calculations it can be proved that
\begin{eqnarray}
H_{I}=H_{S}+H_{T},\qquad H_{S}=\beta V_{S},\qquad H_{T}=i\gamma
V_{T}.
\end{eqnarray}
Here $V_{S}$, $V_{T}$ are functions of $x$. The meaning of $V_{S}$
is transparent. It is an external scalar potential. The tensor part
of the interaction Hamiltonian $H_{T}$ can be considered as
particular case ($A_{\mu}=(A_0(x_1),\vec{0})$) of the more general
tensor Hamiltonian
\begin{eqnarray}
  H_{T}&=&gi\gamma_0\sigma_{\mu\nu}F_{\mu\nu}, \\
  \sigma_{\mu\nu}&=&\gamma_\mu\gamma_\nu-\gamma_\nu\gamma_\mu, \\
  F_{\mu\nu}&=&\frac{\partial A_{\nu}}{\partial x_\mu}-\frac{\partial A_\mu}{\partial
  x_\nu},
\end{eqnarray}
where $\gamma_\mu$ are the Dirac matrices, $A_{\mu}$ is an external
electromagnetic field and $g$ is a coupling constant proportional to
magnetic moment of the particle.

Such Hamiltonian describes the interaction of a  spin-$1/2$ neutral
massive particle (e.g. neutron) with an external electrostatic
field.

From above  consideration we conclude that the tensor potential
$V_{T}$ has meaning  of the electrical strength
\begin{eqnarray}
V_{T}\sim E_{x}=-\frac{\partial A_{0}}{\partial x}.
\end{eqnarray}

\begin{figure}[]
\centering \includegraphics[height=5cm]{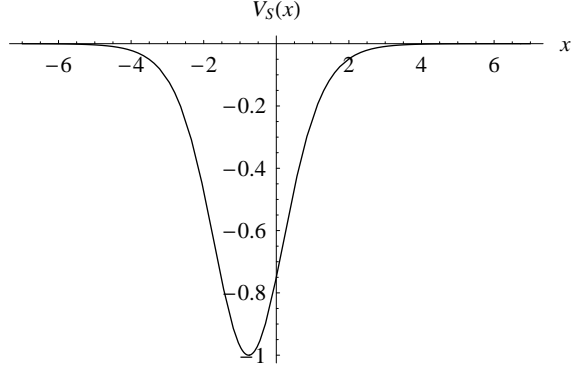} \caption{The scalar
part of the external potential $V_I(x)=V_S(x)+V_T(x)$:
$V_S(x)=-2k^2/(m+\lambda\cosh (2kx+2\phi))$,
$k=\sqrt{m^2-\lambda^2}$, $m=1$, $\lambda=1/2$.} \label{f1}
\end{figure}

\section{Some special choice of parameters}\noindent

\textbf{Case 1.} Let $\lambda_1=\lambda_2=\lambda$,
$\phi_1=\phi_2=\phi$. \noindent In this case $V_{T}=0$
\begin{eqnarray}
V_{S}(x)=-\frac{2k^2}{m+\lambda\cosh{(2kx+2\phi)}},\qquad
k=\sqrt{m^2-\lambda^2}.
\end{eqnarray}
\noindent This result coincides  with the result of \cite{Noga2} at
$\phi=[\ln{(m+k)/(m-k)}]/4.$ \\The function $V_S(x)$ is shown for
fixed  values  $m=1$ and
$\lambda=1/2$ in Figure  \ref{f1}.\\
\begin{figure}[h]
\centering \includegraphics[height=5cm]{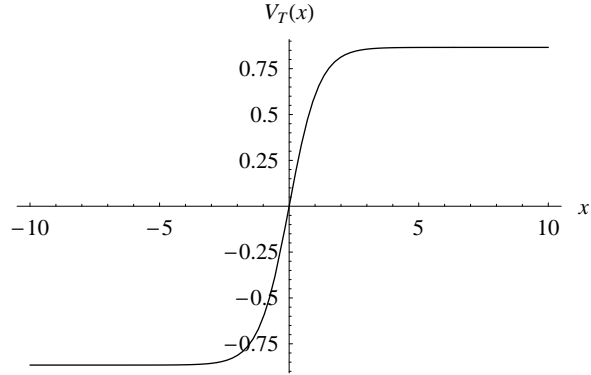} \caption{The tensor
part of the $V_I(x)$: $V_T(x)=k_2\tanh(k_2x)$,
$k_2=\sqrt{m^2-\lambda^2_2}$, $m=1$, $\lambda_2=1/2$.} \label{f2}
\end{figure}
\begin{figure}[b] \centering
\includegraphics[height=5cm]{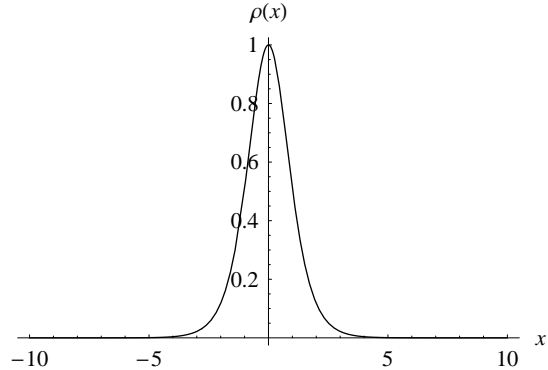} \caption{The charge
distribution $\rho(x)=1/\cosh^2(k_2x)$,
$k_2=\sqrt{m^2-\lambda^2_2}$,  $m=1$, $\lambda_2=1/2$.} \label{f3}
\end{figure}

\textbf{Case 2}. Let $\lambda_1=m$, $\phi_1=\phi_2=0.$ In this case
\begin{eqnarray}
V_{S}=-\lambda_2-m=const
\end{eqnarray}
\noindent which  leads to the redefinition of the particle mass
\begin{eqnarray}
m\longrightarrow-\lambda_2.
\end{eqnarray}
The tensor potential (electrical strength) is
\begin{eqnarray}
\label{VT}
  V_{T}(x)=k_2\tanh{(k_2x)}\sim E_x.
\end{eqnarray}
A pictorial representation of the potential $V_T(x)$  for fixed
values $m=1$ and $\lambda_2=1/2$ is given in Figure \ref{f2}.
\\Such
electrical field is produced by the charge distribution
\begin{eqnarray}
\rho(x)\sim\frac{1}{\cosh^2{(k_2x)}}.
\end{eqnarray}
The charge distribution $\rho$ for values $m=1$ and $\lambda_2=1/2$
is shown  in Figure \ref{f3}. \noindent\\
 It is easy to check  that both $V_{S}$ and $V_{T}$
are transparent potentials.

\section{Simple chain of the Darboux---Crum transformations}

As the next step consider a chain of the Darboux---Crum
transformations for the four-component Dirac equation (\ref{1}). Let
us put in expressions (\ref{10}), (\ref{11}) $\phi_1=\phi_2=0$,
$k_1=0$, $k_2=k$, $\lambda_2=\varepsilon_1=\sqrt{m^2-k^2}.$ This
leads to the Hamiltonian $H_1$ of the following form
$H_1=-i\alpha_1\partial_x-\varepsilon_1\beta+i\gamma k\tanh{(kx)}.$

Consider the matrix
\begin{eqnarray}
\label{71}
  u^{(1)}=a^{(1)}I+b^{(1)}\alpha_1+c^{(1)}\beta+d^{(1)}\gamma,
\end{eqnarray}
where
\begin{eqnarray}
\label{81} a^{(1)}&=&(\mu_1^{(1)}+\mu_3^{(1)})/2,\qquad
b^{(1)}=(\mu_2^{(1)}+\mu_4^{(1)})/2,\\
 c^{(1)}&=&(\mu_1^{(1)}-\mu_3^{(1)})/2,\qquad
 d^{(1)}=(\mu_4^{(1)}-\mu_2^{(1)})/2,\nonumber
\end{eqnarray}
\begin{eqnarray}
\label{101}
 & &\mu_1^{(1)}=\cosh{(kx)},\qquad
\mu_2^{(1)}=\frac{-3ik\sinh{(kx)}\cosh{(kx)}}{\varepsilon_2-\varepsilon_1},\\
&&\mu_3^{(1)}=\cosh^2{(kx)},\qquad\mu_4^{(1)}=0,\qquad
\varepsilon_2=\sqrt{m^2-4k^2}.
\end{eqnarray}
The matrix (\ref{71}) obey the equation
\begin{eqnarray}
  H_1u_1&=&u_1\Lambda_1,\qquad  \Lambda_1=\frac{1}{2}\left[(I+\beta)\lambda^{(1)}_1+(I-\beta)\lambda^{(1)}_2\right],\\
  \lambda^{(1)}_1&=&-\varepsilon_1,\qquad
  \lambda^{(1)}_2=\varepsilon_2.
\end{eqnarray}
Consider the intertwining relation
\begin{eqnarray}
 \label{21} L_1H_1=H_2L_1,
\end{eqnarray}
where
\begin{eqnarray}
\label{31} L=\partial_x-u^{(1)}_x u^{(1)^{-1}}.
\end{eqnarray}
Then
\begin{eqnarray}
  H_2-H_1=-i[\alpha_1,u^{(1)}_xu^{(1)^{-1}}]=\beta(\varepsilon_1-\varepsilon_2)+i\gamma
  k\tanh{(kx)}
\end{eqnarray}
from which it follows
\begin{eqnarray}
  H_2&=&-i\alpha_1\partial_x-\varepsilon_2\beta+i\gamma2k\tanh{(kx)}.
\end{eqnarray}\begin{figure}[h]
\centering \includegraphics[height=5cm]{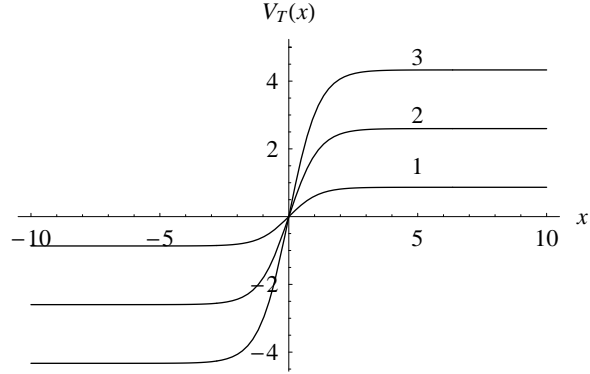} \caption{The
tensor part of the interaction  Hamiltonian (\ref{33}):
$V_T(x)=nk\tanh(kx)$, $k=\sqrt{m^2-\lambda^2_2}$, $m=1$,
$\lambda_2=1/2$ (line 1: $n=1$; line 2: $n=3$; line 3: $n=5$)}.
\label{f2n}
\end{figure}

Let \begin{eqnarray}
  \label{33} H_n&=&-i\alpha_1\partial_x-\varepsilon_n\beta+i\gamma nk\tanh{(kx)},\qquad
  \varepsilon_n=\sqrt{m^2-n^2k^2}
\end{eqnarray}
and
\begin{eqnarray}
\label{7n}
  u^{(n)}=a^{(n)}I+b^{(n)}\alpha_1+c^{(n)}\beta+d^{(n)}\gamma,
\end{eqnarray}
where
\begin{eqnarray}
\label{8n} a^{(n)}&=&(\mu_1^{(n)}+\mu_3^{(n)})/2,\qquad
b^{(n)}=(\mu_2^{(n)}+\mu_4^{(n)})/2,\\
c^{(n)}&=&(\mu_1^{(n)}-\mu_3^{(n)})/2,\qquad
d^{(n)}=(\mu_4^{(n)}-\mu_2^{(n)})/2,\nonumber
\end{eqnarray}
\begin{eqnarray}
\label{10n}
 & &\mu_1^{(n)}=\cosh^n{(kx)},\quad
\mu_2^{(1)}=\frac{-i(2n+1)k\sinh{(kx)}\cosh^{(n-1)}{(kx)}}{\varepsilon_{n+1}-\varepsilon_n},\\
&&\mu_3^{(1)}=\cosh^{(n+1)}{(kx)},\quad\mu_4^{(1)}=0.
\end{eqnarray}
Then $u_n$ obey the equation
\begin{eqnarray}
  H_nu_n&=&u_n\Lambda_n,\quad  \Lambda_n=\frac{1}{2}\left[(I+\beta)\lambda^{(n)}_1+(I-\beta)\lambda^{(n)}_2\right],\\
  \lambda^{(n)}_1&=&-\varepsilon_n,\quad
  \lambda^{(n)}=\varepsilon_{n+1}.
\end{eqnarray}
The solutions of the equation
\begin{eqnarray}
  H_{n+1}\psi_s&=&E_s\psi_s
\end{eqnarray}
corresponding the discrete levels are of the following form:
\begin{eqnarray}
 \label{40} \phi_s&=&(\phi_{s1},\phi_{s2},\phi_{s3},\phi_{s4})^t, \\
  \phi_{s1}&=&AP_n^{s}\{\tanh{(kx)}\},\quad \phi_{s2}=BP_n^s\{\tanh{(kx)}\}, \nonumber\\
  \phi_{s3}&=&CP_{n+1}^s\{\tanh{(kx)}\},\quad \phi_{s4}=DP_{n+1}^s\{\tanh{(kx)}\},\nonumber\\
\frac{A}{D}&=&\frac{B}{C}=\frac{k(n+s+1)}{\varepsilon_{n+1}+E_s},\nonumber\\
E_{s}&=&sign(s)\varepsilon_s,\quad
\varepsilon_s=\sqrt{m^2-s^2k^2},\\
s&=&-n,-n+1,...,-2,-1,1,2,...,n,n+1,
\end{eqnarray}
where $P_n^s$ are associated Legendre polynomials \cite{Rygik}, $D$
and $C$ are arbitrary nonzero numbers.\\
The solutions of the equation
\begin{eqnarray}
  H_{n+1}\psi&=&E\psi,\qquad |E|>m,
\end{eqnarray}
corresponding to the continuous spectrum can be obtained from
(\ref{40}) by the substitution
\begin{equation}
  P^s_{n(n+1)}\{\tanh{(kx)}\}\Rightarrow P^{\mu}_{n(n+1)}\{\tanh{(kx)}\},\quad
  \mu=\pm ip/k,\quad p=\sqrt{E^2-m^2},
\end{equation}
\begin{equation}
  P_n^{(\pm ip/k)}\{\tanh{(kx)}\}=\frac{\exp{(\pm
  ipx)}}{\Gamma(1\mp ip/k)}\cdot F\left(-n;n+1;1\mp \frac{ip}{k};\frac{1-\tanh (kx)}{2}\right).
\end{equation}
From the last expression  it is evident  that ``potentials'' of the
form $-\beta\varepsilon_n+i\gamma n\tanh{(kx)}$ are transparent.

\section{Exactly solvable periodic potentials}
Starting with exactly solvable potential for full real axis one can
construct the exactly solvable periodic potentials in the manner
described in the papers \cite{per2,Eurjphys,per3}.

It is well known that the continuous spectrum of such potentials
have a band structure. For the investigation of this structure it is
necessary to construct so-called {\it Lyapunov function} $D(E)$ (see
for details \cite{per2,Eurjphys,per3}).

The values of energy $E$ which obey the equation $|D(E)|>2$ belong
to the forbidden bands. All others belong to the allowed bands.

Omitting simple but cumbersome calculations  we present here only
final result for the some lowest values of the $n$:

 \begin{equation}
D_n(E)=2[\cos(2pa)A_n(p,t_1)-(kt_1/p)\sin(2pa)]B_n(p,t_1),
\end{equation}
 \begin{equation}t_1=\tanh(ka),\qquad a=T/2,
\end{equation}
where $T$ is the parameter of periodization
\cite{per2,Eurjphys,per3}. In general case (of the arbitrary $n$)
\begin{equation}
  A_n=\frac{Im \Phi_n(\nu,t_1)}{\nu},\qquad B_n=\frac{Re \Phi_n(\nu,t_1)}{t_1},\end{equation}
  \begin{equation}\Phi_n= (n-i\nu)
\cdot F\left(-n+1,n;1-i\nu;\frac{1-t_1}{2}\right)\cdot
F\left(-n,n+1;1+i\nu;\frac{1+t_1}{2}\right),\end{equation}
  $$\nu=p/k.$$
\begin{figure}[ph] \centering
\includegraphics[height=5cm]{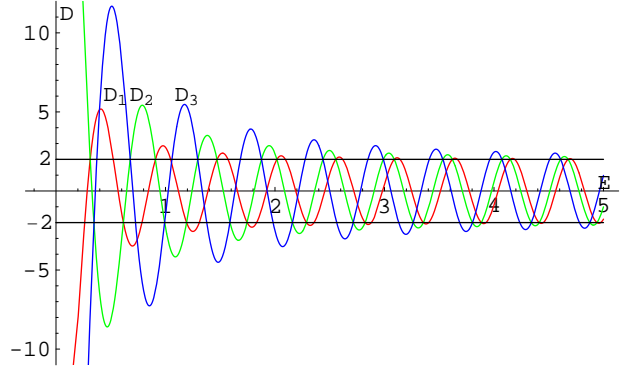}
\caption{Lyapunov functions $D_n,$ $n=\overline{1,3}$, with $ka=6$.}
\label{Figure7}
\end{figure}
In the case $ka>>1$ ($1-t_1<<1$) 1), it is readily found from
(\ref{DnE}) and (\ref{t1})
\begin{equation}
D_n(E)=2\frac{\cos(\phi_n-\arctan(nk/p))}{\cos(\arctan(nk/p))},
\end{equation}
\begin{equation}\label{phin}
    \phi_n=2[pa+\sum_{s=1}^n\arctan(sk/p)].
\end{equation}
The solution of inequality $|D_n(E)|\geq 2$ is trivial in this case.
\\Lyapunov functions $D_n(E)$ with
$n=\overline{1,3}$ are presented in Fig. \ref{Figure7} for $ka=6$.
\section{Discussion}

We show that the Darboux transformation for the free one-dimensional
four-compo\-nent Dirac Hamiltonian generates the interaction that can
takes  place in the real world.

The scalar potential  arises in the theory of nuclear-nuclear
interactions \cite{Sucher,Ignatovish} when sigma-exchange is taken
into account \cite{Partovi,Donoghue} and can be considered as a part
of the nuclear potential in the channeling theory
\cite{YFN,Fedorov}.

The interaction of a neutral massive particle with the spin $1/2$
(e.g.neutron) with an external electrostatic field generated by the
Darboux transformation can be interpreted as Schwinger interaction
of neutron with atoms of the lattice in its channeling in the
nonmagnetic crystals
\cite{YFN,Kadmenski,Chan,Ryab,channeling2004,channeling2005,channeling}.

The origin of the Schwinger interaction  is due to relativistic
effect of appearance of the lattice magnetic field in the neutron
rest frame, when it passes through the electrical field of the
crystal. Due to proper magnetic moment of neutron the appearance of
such field leads to magnetic interaction, which allows to order and
orientate movement of neutron as in the case of channeling in
nonmagnetic crystal \cite{YFN,Fedorov,Lap2,EMD,SHVINGER}.

Really, in the coordinate system connected with the neutron moving
in the electrical field of the crystal the magnetic field
$\vec{H}=[\vec{E},\vec{v}]$ appears. In the case of the plane
channel the value of the $\vec{E}$ averaged by the area of
elementary cell goes to zero. As a result, the mean value of the
$\vec{H}$ obtains its maximum near the plane. In the middle of two
planes $\vec{H}$ goes to zero, changes the sign and increases to
maximum value near the next plane of the channel (see Figures 2 and
4).

Moving across the plane the effective field $\vec{H}$, due to the
changing of the sign of the electrical field vector $\vec{E}$,
change the sign and its space behaviour repeating the behaviour in
the previous channel (Figure 5). \begin{figure}[h] \centering
\includegraphics[height=5cm]{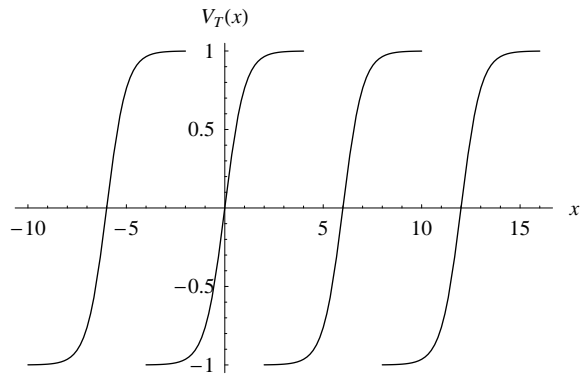} \caption{Periodic
continuation of the potential \eqref{VT} for $k_2=1$.}
\label{periodicf2b}
\end{figure}

In accordance with the accepted terminology, the considered
neutron-cell interaction may be characterized as \emph{coherent
Schwinger interaction} and is the generalization of well known
\emph{neutron-atom Schwinger interaction} \cite{YFN}.

\section*{Acknowledgments}

The author is grateful to Prof. A. V. Tarasov for fruitful
discussions and  Dr. S. R. Gevorkyan who have read the paper and
made useful comments. I am grateful to Joint Institute for Nuclear
Research (Dubna, Moscow region, Russia) for hospitality during this
work. The work was supported in part by the grant of Dinastiya
Foundation for Noncommercial Programs and Moscow International
Center of Fundamental Physics.

\end{document}